\documentclass[letterpaper,11pt,onecolumn,accepted=2025-11-17]{quantumarticle}
\pdfoutput=1
\usepackage[utf8]{inputenc}
\usepackage[english]{babel}
\usepackage[T1]{fontenc}
\usepackage{hyperref}

\usepackage{amsmath,amsthm,amsfonts,amssymb,braket}
\usepackage[all]{xy}
\usepackage{fullpage}
\usepackage{mathtools}
\usepackage[numbers]{natbib}

\newtheorem{proposition-definition}[lemma]{Proposition-Definition}
\theoremstyle{definition}

\newcommand{\CC}{{\mathbb{C}}}

\newcommand{\ZZ}{{\mathbb{Z}}}

\newcommand{\calA}{{\mathcal{A}}}
\newcommand{\calB}{{\mathcal{B}}}

\newcommand{\one}{{\mathbf{1}}}

\usepackage{tikz}
\usepackage{lipsum}

\newcommand{\SU}{{\mathsf{SU}}}

\newcommand{\U}{{\mathsf{U}}}

\newcommand{\uone}{{\mathsf{U}(1)}}
\newcommand{\swap}{{\mathsf{S}}}

\DeclareMathOperator{\poly}{poly}
\DeclareMathOperator{\polylog}{polylog}

\newcommand{\ii}{{\mathbf{i}}}

\DeclareMathOperator{\Tr}{{Tr}}

\DeclareMathOperator*{\Supp}{{Supp}}

\DeclareMathOperator{\avg}{{\mathbb{E}}}
\DeclareMathOperator*{\Mat}{{\mathsf{Mat}}}

\DeclarePairedDelimiter{\norm}{\lVert}{\rVert}
\DeclarePairedDelimiter{\abs}{|}{|}

\begin{document}

\title{Short remarks on shallow unitary circuits}

\author{Jeongwan Haah}
\affiliation{Leinweber Institute for Theoretical Physics, Stanford University,\\
Google Quantum AI}
\orcid{0000-0002-1087-6853}
\date{}

\maketitle 

\begin{abstract}
    (i) We point out that every local unitary circuit
    of depth smaller than the linear system size
    is easily distinguished from a global Haar random unitary
    if there is a conserved quantity that is a sum of local operators.
    This is always the case with a continuous onsite symmetry
    or with a local energy conservation law.
    
    (ii) We explain a simple algorithm for a formulation of 
    the shallow unitary circuit learning problem
    and relate it to an open question on strictly locality-preserving unitaries (quantum cellular automata).

    (iii) We show that any translation-invariant quantum cellular automaton
    in $D$-dimensional lattice of volume~$V$
    can be implemented using only $O(V)$ local gates in a staircase fashion 
    using invertible subalgebra pumping.
\end{abstract}

%\maketitle

\section{Symmetric shallow circuit $\neq$ symmetric global random unitary}

Recently it has been shown~\cite{LaRacuente2024,Schuster2024}
that an ensemble of random local quantum circuits $U$
on a one-dimensional chain of $n$ qubits of depth~$k \polylog(n k)$
is indistinguishable from the Haar random global unitary on $n$ qubits
if we query $U$ no more than $k$ times.%
\footnote{%
    The indistinguishability can also be established under a cryptographic assumption 
    with a milder restriction on the number of queries to the unitary~\cite{Schuster2024}.
}
Such an ensemble of unitaries is called a \emph{unitary $k$-design}.
The existence of a shallow unitary design is rather surprising
since the depth is much smaller than the linear system size~$n$,
so the absolute lightcone of an observable at one end of the chain does not even reach the other end.
Nonetheless, the ensemble contains sufficient randomness 
which prohibits any predetermined correlation function
from probing the difference between Haar random global unitaries and shallow local circuits.

Here we point out that a certain symmetry condition
rules out all shallow unitary $k$-designs for any $k \ge 1$
because there is a useful predetermined autocorrelation function.
This extends the list of conditions under which shallow random unitary designs are impossible.
Previously it was known~\cite{Hearth2023} that
a brickwork random circuit of depth~$o(L^2)$ 
in Euclidean lattice of linear size~$L$ 
that conserves $U(1)$ charge
remains distinguishable from the global $U(1)$-symmetric Haar random unitaries;
it was also known~\cite{Schuster2024} that
any ensemble of shallow circuits is distinguishable from Haar random unitaries
if there is time-reversal symmetry or
if we can access the inverse of the unitary.
In essence, our argument is that
if we have a local charge density 
whose integral (sum) is conserved,
then a shallow circuit and a symmetric global Haar random unitary are very different
since local unitaries may spread charges only to a small distance
while a symmetric Haar random unitary spreads charges across the whole system.%
\footnote{See~\cite{Khemani2017,Rakovszky2017} for related results.}
This summary is good enough as a take-away,
but we need an extra ingredient for a sound protocol.

We work with a one-dimensional array of $n$ qudits~$\CC^p$.
So, the total Hilbert space is $(\CC^p)^{\otimes n}$.
Consider a symmetry (generator)
\begin{equation}
    Q = \sum_{i=1}^n Q_i
\end{equation}
where $Q_i$ is a nonzero hermitian operator supported on a qudit~$i$.
We assume that every $Q_i$ has an eigenvalue zero.%
\footnote{%
    One may further require that each $Q_i$ have an integer eigenvalue spectrum
    so that $e^{\ii \theta} \mapsto e^{\ii \theta Q}$ is a representation of~$\uone$,
    but this condition is not used in our argument.
}
This condition is trivially fulfilled by adding a real multiple of identity to~$Q_i$.
Let $\swap_{i,j} = \swap_{j,i}$ be the swap operator that interchanges qudit~$i$ and~$j$.
For simplicity we assume that 
$    Q_i = \swap_{i,j} Q_j \swap_{i,j}^\dagger $.
The symmetric unitary group of interest 
consists of all $U \in \U(p^n)$ that commutes with~$Q$.
A symmetric local unitary circuit is a composition of layers of nonoverlapping
nearest-neighbor 2-qudit unitaries, \emph{each of which commutes with~$Q$}.
Of course, any symmetric local unitary circuit belongs to the symmetric unitary group.%
\footnote{%
    The converse is false;
    there exists a symmetric unitary 
    that is not expressible by any symmetric local unitary circuit~\cite{Marvian2020}.
    This is however unimportant since we focus on a lower bound on the depth,
    which is still meaningful since symmetric local unitary circuits with sufficient depths
    can form a symmetric unitary $k$-design
    for~$k$ that grows with the system size~\cite{Hearth2023}.
}
The depth is the number of layers.
This is a standard setting in the context of symmetry-protected topological phases~\cite{Chen2010}.

\begin{figure}
    \centering
    \includegraphics[width=\textwidth, trim={2ex 90ex 5ex 20ex}, clip]{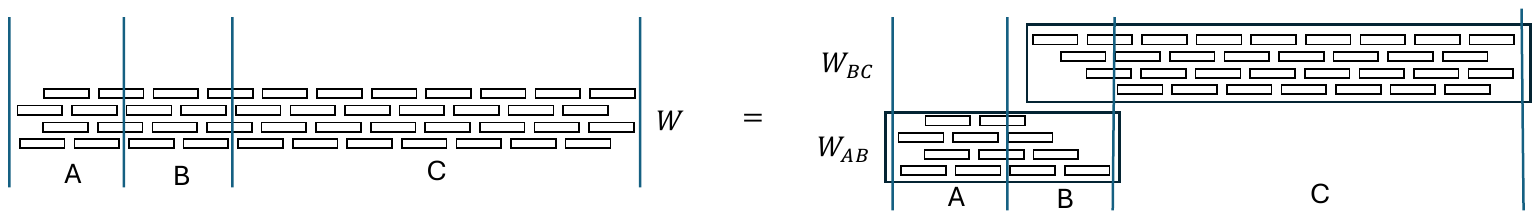}
    \caption{
        Any circuit~$W$ of depth smaller than the linear system size is a product of~$W_{AB}$ and $W_{BC}$.
    }
    \label{fig:circuit}
\end{figure}

The distinguishing protocol is the following.
We partition the system into three disjoint intervals $A,B,C$ 
such that $\frac 1 5 n \le \abs A = \abs B \le \frac 1 4 n \le \abs C $.
Define $Q_S = \sum_{i \in S} Q_i$ for any subset~$S$ of qudits.
We initialize a state~$\rho_0$ 
so that $\Tr(Q_A \rho_0) = q > 0$ but $\Tr(Q_B \rho_0) = 0 = \Tr(Q_C \rho_0)$.
Then, we apply a symmetric Haar random unitary~$U_{AB}$ supported on~$AB$.
We next apply a given instance~$W$ from an unknown ensemble of symmetric unitaries.
We finally measure~$Q_A$.

Suppose that $W$ is drawn from an ensemble of symmetric local circuits of depth $\le \frac 1 {100} n$.
Since $A$ and $B$ are sufficiently big ($\ge \frac 1 5 n$),
the instance~$W$ can be written as a product of two supergates: $W = W_{BC} W_{AB}$. 
See Figure~\ref{fig:circuit}.
Note that $W_{AB}$ commutes with~$Q_{AB}$ and $W_{BC}$ commutes with~$Q_{BC}$. 
Due to the randomization~$U_{AB}$ just before~$W$,
the component~$W_{AB}$ does nothing statistically;
the probability distributions of $U_{AB}$ and of~$W_{AB} U_{AB}$ are identical.
Therefore, immediately after $U_{AB}$, the partial sum~$Q_A$ has an expectation value
$\overline{Q_A} = \frac 1 2 q$ (over the random choices of~$U_{AB}$)
because the distribution of~$U_{AB}$ is invariant under the swap of~$A$ and~$B$.
The final measurement of~$Q_A$ is unaffected by~$W_{BC}$.

On the other hand, 
if $W = U_{ABC}$ is drawn from the symmetric Haar distribution over the entire system,
then the randomization $U_{AB}$ of the protocol does nothing statistically;
the probability distribution of~$U_{ABC}$ and of~$U_{ABC}U_{AB}$ are identical.
The final measurement gives an expectation value~$\overline{Q_A} = \frac{\abs A}{n} q \le \frac 1 4 q$.
The measured $Q_A$ is a random variable valued in~$[-q_0, q_0]$ where $q_0 = \abs A \norm{Q_i}$.
If we set $q$ to be of order~$\abs A \norm{Q_i}$,
the expectation values of~$Q_A$ in the two scenarios are 
comparable to or larger than the statistical fluctuations,
where the latter will become small, after a few repetitions,
compared to the difference $\frac 1 2 q - \frac 1 4 q = \frac 1 4 q$.
We can therefore distinguish whether $W$ is from the symmetric Haar random ensemble 
or from a shallow symmetric local circuit with confidence~$1-\delta$ 
after $O(\log \frac 1 \delta)$ repetitions of the protocol.
Each run queries the given $W$ once.

Since our protocol is explicit,
it is at least as strong a notion of distinguishability
as the additive error for unitary designs,
which is more of an information-theoretic measure.
Under the latter notion
where two mixed unitary channels are sufficiently far apart in the diamond distance,
input states and output measurements for the task of distinguishing channels
may be implicit and difficult to find.

This completes the analysis of the protocol.
In conclusion, for any ensemble of symmetric local circuits
to form an approximate symmetric unitary $k$-design with $k \ge 1$ and a small constant additive error,
the depth of the circuit must be at least linear in the linear system size.
Some comments are below.

The randomization by~$U_{AB}$ is important to make the protocol sound.
If we omitted it, 
the measurement of~$Q_A$ may give a misleading answer for the distinguishing task
because symmetric circuits can conspire to transport an abelian charge 
from anywhere to anywhere instantaneously by creating and annihilating charge-anticharge pairs.%
\footnote{%
    Note that $U_{AB}$ is the only component 
    for which the implementation gate complexity as a function of~$n$ 
    might be high;
    other components beyond querying a blackbox~$W$
    are 
    preparation of eigenstates of~$Q_i$
    and measurement of~$Q_i$,
    both of which are local operations.
    Since the protocol is to measure 
    $\overline{Q_A} = \avg_{U_{AB},W}\Tr(Q_A W U_{AB} \rho_0 U_{AB}^\dagger W^\dagger)$
    with a small additive error,
    we only need $U_{AB}$ to be drawn from an approximate symmetric unitary $1$-design.
    It appears that the lower bound on the convergence rate of $U(1)$-symmetric circuits 
    to $U(1)$-symmetric approximate designs is largely an open problem.
}
\begin{figure}
    \centering
    \includegraphics[width=0.7\textwidth, trim = {0ex 95ex 90ex 22ex}, clip]{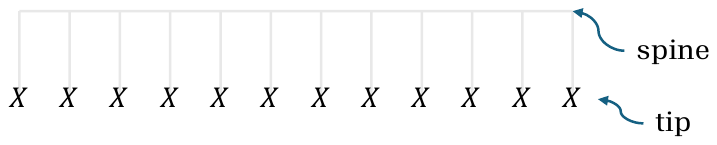}
    \caption{Comb connectivity. 
        A qudit is placed on every tip and spine vertex.
        A two-qudit gate may only act on an edge.
        A discrete symmetry acts on the tip qudits.
    }
    \label{fig:comb}
\end{figure}

The existence of a conserved total charge~$Q$ is implied by a continuous symmetry,
but that is not the only source.
The conserved quantity $Q$ can be a local Hamiltonian itself,
in which case our protocol is simply to measure the energy density.%
\footnote{Some other anticoncentration measures~\cite{Tirrito2024} 
    were numerically observed to grow qualitatively differently 
    depending on whether the dynamics conserves energy.}
Even with a discrete symmetry, a conserved quantity that is a sum of local operators
may appear in the following ``dilute symmetry'' scenario.
Suppose that we have a \emph{comb}\footnote{I thank Ehud Altman for suggesting a comb to explain this dilute symmetry.}
connectivity graph as shown in Figure~\ref{fig:comb}.
A qudit is placed on every vertex, and 2-qudit local unitaries are allowed only over the edges.
Label each qudit by~$(i,\text{tip})$ or~$(i,\text{spine})$.
We impose a symmetry
by a tensor product operator~$Q' = \prod_i X_{(i,\text{tip})}$.
Here the symbol~$X$ is reminiscent of Pauli~$X$ as in the onsite $\ZZ_2$ symmetry on qubits.
Then, every local unitary gate sees at most one tensor factor of~$Q'$
and hence $Q = \sum_i X_{(i,\text{tip})} + X_{(i,\text{tip})}^\dagger$,
in addition to~$Q'$,
is preserved by any symmetric local circuit dynamics.
A symmetric global unitary is assumed to preserve~$Q'$ but not necessarily~$Q$.
This relaxed symmetry condition only widens the gap in the expectation values of~$Q_A$
and the protocol above still disambiguates 
shallow symmetric local circuits from global symmetric Haar random unitaries.

We have assumed that the sizes of subsystems $\abs A$ and $\abs B$ are equal,
that the total system is a one-dimensional array of qudits,
and that all the local operators~$Q_i$ are the same up to relabeling of the qudits.
These assumptions are for simplicity in presentation and are unimportant.
The subsystem~$A$ is where we put some known value of charge.
If $A$ and $B$ are comparable in size,
then a symmetric Haar random unitary~$U_{AB}$ on~$AB$ 
will evenly distribute the charge over~$AB$
and $A$ will retain a significant fraction of the initial charge.
In the symmetric shallow circuit case, the measurement of~$Q_A$ will reveal this
large fraction of the initial charge,
but in the global symmetric Haar random case, the measured value of~$Q_A$ 
will be suppressed by a factor of the overall system size.
It remains to investigate the case of discrete symmetries
represented by product operators that are not necessarily dilute.

We would like to conclude this section with a rather philosophical remark.
Physics 
--- an endeavor to describe complex phenomena by simple principles ---
is possible because the physical principles are testable 
despite of all the inaccuracies in our measurements and descriptions.
If those inaccuracies resemble, hypothetically, the shallow 
random circuit that appears in~\cite{LaRacuente2024,Schuster2024},
then everything would look completely dull
that we would not be able to speak of any phase of matter.
But, fundamentally,
energies and symmetry charges do exist and are conserved,
based on which we have separated any shallow circuits from a global featureless unitary.

\section{Shallow circuit learning}

It was asked in~\cite{Huang2024}
how to output a description of a shallow unitary circuit~$U$ in a $D$-dimensional lattice~$\Lambda$ of qudits
given oracle access to~$U$, while minimizing the number of queries to~$U$ 
as a function of the system size and other parameters.
The answer in~\cite{Huang2024} 
does not give a description of~$U$ 
but instead a circuit on a doubled system that implements~$U \otimes U^\dagger$.
A well-known identity~\cite{Arrighi2011} was used in~\cite{Huang2024}:
\begin{align}
    U \otimes U^\dagger &= (U \otimes \one) \swap (U^\dagger \otimes \one) \swap
\end{align}
where $\swap$ is an operator that swaps the entire system~$\Lambda$ and its copy~$\Lambda'$.
Since the underlying operator algebra is a tensor product of those of individual qudits,
we have
\begin{equation}
    \swap = \prod_{i} \swap_{i,i'}
\end{equation}
where $\swap_{i,i'}$ is the swap operator between $i \in \Lambda$ and its copy~$i' \in \Lambda'$. 
It follows~\cite{Arrighi2011} that,
\begin{equation}
    U \otimes U^\dagger = \left[ \prod_{i} (U \otimes \one) \swap_{i,i'} (U^\dagger \otimes \one) \right] \prod_i \swap_{i,i'} \, . \label{eq:uswap}
\end{equation}
Since $U$ is shallow, the conjugation by~$U$ maps a local operator to a potentially slightly bigger operator nearby.
We will refer to this blow up in support as the {\bf spread} of~$U$.
Therefore, $U \otimes U^\dagger$ is a shallow depth quantum circuit 
consisting of $\swap_{i,i'}$ and their conjugates.
It is now clear that the circuit learning problem can be solved, if we are fine with using an auxiliary system,
by figuring out the conjugates of~$\swap_{i,i'}$.
Note that since $\swap_{i,i'}$ for different $i$'s are commuting,
the ordering of the factors in the product of~\eqref{eq:uswap} is immaterial.

For the last step, we recall a simple strategy that appeared in~\cite{Haah2021}
where it was necessary to estimate $\Tr(UXU^\dagger Y)$ for various local (Pauli) operators~$X$ and $Y$
given query access to~$U$.
Take any density matrix~$\rho \succeq 0$ that, as an operator, is supported on some small set of qudits.
This mixed state~$\rho$ can be prepared 
by some local quantum channel on its support and some other channel 
that prepares all the qudits elsewhere in a random product state.
The conjugate~$U \rho U^\dagger$ is a density operator supported on a slightly bigger set of qudits,
and usual state tomography can determine $U \rho U^\dagger$ whose complexity does not depend on the system size
but only on the depth of~$U$.%
\footnote{
    If $U$ is a shallow circuit consisting of gates from a discrete set
    ({\it e.g.}, Clifford $+$ T),
    then the complexity depends only on the depth.
    If the gates comprising $U$ are from, say, $\SU(4)$,
    then $U\rho U^\dagger$ can only be determined up to some error with some high confidence,
    where the accuracy and confidence must be boosted for various global requirements,
    and system size dependence may become necessary.
    If the tomography has $\sim \epsilon^{-2} \log(1/\delta)$ sample complexity
    for accuracy~$\epsilon$ in some state distance and confidence~$1 - \delta$,
    then one would want to consume $\sim n^2 \epsilon^{-2} \log(n/\delta)$ samples of $U \rho U^\dagger$
    to ensure that the global accuracy of~$U \otimes U^\dagger$ is at most~$\epsilon$ with confidence~$1-\delta$.
}
We can use this to learn~$(U \otimes \one) \swap_{i,i'} (U^\dagger \otimes \one)$
even though the unitary $\swap_{i,i'}$ is \emph{not} a density operator.
Write $\swap_{i,i'} = \frac 1 p \sum_P P_i \otimes P^\dagger_{i'}$ 
as a linear combination of (generalized) Pauli operators,
where $p$ is the local qudit dimension.
Clearly, it suffices to learn $U P U^\dagger$ for every (generalized) Pauli operator~$P$,
but since $P \mapsto U P U^\dagger$ is an algebra homomorphism,
it suffices to learn just two operators~$U X U^\dagger$ and~$U Z U^\dagger$
where $X = \sum_{j\in\ZZ/p} \ket{j+1}\!\bra{j}$ 
and $Z = \sum_{j \in \ZZ/p} e^{2\pi \ii j /p} \ket j \!\bra j$.
Any operator~$O = \frac{O + O^\dagger}{2} + \ii \frac{O - O^\dagger}{2\ii}$ 
is a linear combination of two hermitian operators,
and any hermitian operator~$H$ is a linear combination of two normalized density operators:
\begin{equation}
    H = \sum_j \lambda_j \ket j \! \bra j = 
    \alpha
    \left[\frac{1}{\alpha} \sum_{j : \lambda_j > 0} \lambda_j \ket j \! \bra j \right]
    -
    \beta
    \left[\frac{1}{\beta} \sum_{j : \lambda_j < 0} \abs{\lambda_j} \ket j \! \bra j \right]
\end{equation}
where $\alpha = \sum_{j : \lambda_j > 0} \lambda_j$ and $\beta = \sum_{j : \lambda_j < 0} \abs{\lambda_j}$.
Hence, $O$ is a linear combination of four density operators.
It follows that, by multiplying and summing the results of tomography of eight local states,
we infer $(U \otimes \one) \swap_{i,i'} (U^\dagger \otimes \one)$.
The needed accuracy of the tomography is $O(\epsilon/n \poly(p))$ in operator norm
to ensure that the error in the reconstructed~$U \otimes U^\dagger$ is at most~$\epsilon$.
If each site is a qubit, it suffices to determine 
just two density operators~$U \frac{\one + \sigma^x_i}{2} U^\dagger = \frac 1 2 U \sigma^x_i U^\dagger + \frac 1 2 \one$ 
and $U \frac{\one + \sigma^z_i}{2} U^\dagger$ for each~$i$,
which are $U$-evolved Pauli stabilizer states.
The qubit case is also found in~\cite{Huang2024,Wu2025}.

As used in~\cite{Haah2021},
the learning of $U X_i U^\dagger$ for a set of~$i$'s that are 
sufficiently separated can be performed simultaneously.
This is because the tomography of the state $U (\prod_i \rho_i ) U^\dagger$, 
where $\rho_i$ are density matrices whose supports are separated by distance twice the spread of~$U$,
can be performed independently around each~$i$.
This reduces the number of queries to~$U$ by a factor of roughly~$n/v$, 
where $n$ is the total number of qudits and 
$v$ is the number of qudits in a ball of radius equal to the spread of~$U$,
compared to the naive strategy 
where we determine each $U X_{i} U^\dagger$ at a time.

\vspace{1ex} {\bf QCA.} \hspace{1ex}
We used only the condition that $U$ maps a local operator to a local operator in the above algorithm,
and hence $U$ could have been any quantum cellular automaton (QCA) ---
by definition, a QCA~$\alpha$ is an automorphism%
\footnote{%
    If the operator algebra is finite dimensional,
    then the Skolem--Noether theorem says that there exists a unitary~$U$ such that
    $\alpha(X) = U X U^\dagger$ for all operators~$X$.
    Hence, it is safe to regard~$\alpha$ as some global unitary.
}
of an operator algebra
such that there exists a constant~$r > 0$ such that
for any single-site operator~$X$, the image~$\alpha(X)$ is supported on the $r$-neighborhood of the support of~$X$.
The minimal constant~$r$ here is the spread of a QCA.
If a circuit has depth~$d$, then we know that its spread 
(of the QCA defined by the conjugation by the circuit)
is~$r = O(d)$.
A primary question in the context of QCA is the converse:
given an automorphism~$\alpha$ of a local operator algebra of spread~$r$,
does there exist a quantum circuit of depth~$d$ that implements~$\alpha$?
If so, how does $d$ depend on the system size and the spread?
If $d$ is a constant independent of the system size,
then the QCA is said to be trivial.
It is quite a widely open problem 
how to determine whether a given QCA,
or an infinite sequence of QCA of a fixed spread on finite systems of diverging sizes,
is trivial.%
\footnote{%
    This is well understood in one dimension with qudits without tails~\cite{Gross2012},
    with tails~\cite{Ranard2020},
    with fermions~\cite{Fidkowski2017}, or with symmetry~\cite{Ma2024};
    and in two dimensions with qudits without tails~\cite{Freedman2019a} (see also \cite{Haah2019});
    and in any dimensions in the translation invariant prime Clifford category~\cite{Haah2025},
    but other situations remain open.
    See the introduction of~\cite{Haah2025} and references therein.
}
The above algorithm outputs a shallow circuit that implements $\alpha \otimes \alpha^{-1}$,
but does not address the QCA triviality problem.
A related question for the circuit learning problem is 
if it is possible to output a circuit (of a similar depth to the input circuit's)
that does not use any ancillas.
Interestingly, a related state learning problem uses an arbitrarily small density of ancillas
to output a circuit that creates a given shallow-circuit-evolved product state~\cite{Landau2024,Kim2024}.

\section{Gate complexity of QCA}

As discussed above, 
a nontrivial QCA is a quantum circuit whose spread is small but depth is large.
For example, a shift QCA in one dimension
can be implemented by a staircase circuit of swap gates on any circle,
where the spread is~$1$, but the depth is the circumference of the circle.
Though it is a deep circuit, the gate count is only the number of sites.
Such a circuit has been considered under the name of sequential quantum circuits~\cite{Chen2023}.
It is shown in~\cite{Chen2023} using a characterization of QCA~\cite{Schumacher2004} 
by generalized Margolus partitioning that
\emph{for any translation invariant%
\footnote{We believe that the assumption of translation invariance is merely a technical convenience,
    but we do not know of a simple enough argument that would cover cases without translation invariance.}
 QCA~$\alpha$ of spread~$r$ 
on a $D$-dimensional system of $n$ qudits~$\CC^p$ with periodic boundary conditions,
there exists a unitary circuit~$U$ consisting of~$O(n)$ gates
such that $\alpha(x) = U x U^\dagger$ for all operator~$x$.}
Hence, in terms of gate count, every QCA is similar to a constant depth quantum circuit.

Here, we present another proof of the theorem of~\cite{Chen2023}
by invertible subalgebra pumping.
Our proof is inductive in the spatial dimension~$D$ where the base case is $D=1$.%
\footnote{We could take $D=0$ as the base case, 
but it would be more instructive to start with $D=1$.}

With $D=1$, a QCA is completely classified~\cite{Gross2012} by a rational index
that captures the ``flow.'' 
If the qudit dimension~$p$ is a prime, the rational index is
any integer (zero, positive, or negative) power of~$p$.
An index~$p^e$ is achieved by a shift QCA that sends every qudit along the positive axis direction by $e$ sites;
if $e < 0$, the shift is to the left.
The spread is~$\abs e$.
A theorem of~\cite{Gross2012} is that if the index is $1 = p^0$,
then the QCA is a constant depth quantum circuit.
The shift QCA of index~$p^1$ is implemented by a staircase of swap gates
as one can check easily.
Hence, the shift QCA of index~$p^e$ has a staircase circuit that is the $\abs e$-fold composition
of the swap-staircase.
So, any QCA~$\alpha$ is a composition of a shift and a quantum circuit of depth depending only on the spread of~$\alpha$,
and the total gate count is $O(n)$ where the big-O notation hides a spread-dependent constant.
If $p$ is a composite number, then we can implement a shift QCA for each prime factor of~$p$.
The gate count is still~$O(n)$.

With $D=2$, one can use a classification result~\cite{Freedman2019a} (see also~\cite{Haah2019}),
but we give a more general argument, which will complete the inductive step of the proof.
We will use a result of~\cite{Haah2022}
that says every QCA in any dimension is basically a shift in disguise.
To explain this, we recall the following definition.
Let $\Lambda_{D-1}$ be a 
% finite%
% \footnote{%
%     In~\cite{Haah2022} we considered infinite lattices~$\ZZ^D$
%     because that was more convenient for presentation, 
%     but, here, since we are going to discuss complexity,
%     it is better to think of a QCA as an infinite sequence 
%     of algebra automorphisms on finite systems of increasing sizes
%     such that the spread is uniformly bounded.
% }
$(D-1)$-dimensional lattice of qudits~$\CC^p$.
An {\bf invertible subalgebra~$\calA$} is a subalgebra of the tensor product algebra
$\Mat(\Lambda_{D-1}) = \bigotimes_{i \in \Lambda_{D-1}} \Mat(p;\CC)_i$
such that the natural multiplication map
\begin{equation}
    \calA \otimes \calB \ni a \otimes b \mapsto ab \in \Mat(\Lambda_{D-1}),
\end{equation}
where $\calB$ is the commutant of~$\calA$ within~$\Mat(\Lambda_{D-1})$,
\emph{has a locality-preserving inverse}.
By locality preserving we mean that there exists $r > 0$ such that
every operator~$x \in \Mat(\Lambda_{D-1})$ can be written as $x = \sum_j a_j b_j$
where $a_j \in \calA$, $b_j \in \calB$, and both $\Supp(a_j)$ and $\Supp(b_j)$ 
are contained in the $r$-neighborhood of~$\Supp(x)$.
The parameter $r$ here is independent of the system size, and is analogous to the spread of a QCA.
In zero dimensions, the local operator algebra~$\Mat(\Lambda_0)$ is a finite dimensional matrix algebra,
and every invertible subalgebra is a tensor factor.
Because of the inverse, the multiplication map is an locality-preserving isomorphism of algebras
and we may write $\calA \otimes \calB \cong \Mat(\Lambda_{D-1})$.

Given an invertible subalgebra~$\calA$ of~$\Mat(\Lambda_{D-1})$,
we construct an associated QCA in $D$ dimensions as follows.
Consider an (infinite) stack of $(D-1)$-dimensional sheets.
Each sheet supports a copy of~$\Mat(\Lambda_{D-1})$.
The full algebra of the stack, $\Mat(\Lambda_D)$, is the (infinite) tensor product of those of the sheets.
To define an algebra endomorphism on~$\Mat(\Lambda_D)$, 
it suffices to specify the images of generators,
which we take to be single-qudit operators.
A single-qudit operator~$x$ is supported on, say, the $k$-th sheet,
and we have a decomposition~$x^{(k)} = \sum_j a_j^{(k)} b_j^{(k)}$ according to the definition
of~$\calA$.
Declare that $x^{(k)} \mapsto \sum_j a_j^{(k)}b_j^{(k+1)}$,
where $b_j$ are sent to the next sheet.
It can be verified~\cite[3.9]{Haah2022} that this is indeed an algebra automorphism
that preserves the locality with a spread bounded by a function of~$r$ 
that appears in the definition of~$\calA$.
We denote by~$\beta_\calA$ the QCA defined from~$\calA$.

A theorem~\cite[3.14]{Haah2022} is that
for every QCA~$\alpha$ of a constant spread~$r$ on~$\Mat(\Lambda_D)$
there exists a shift QCA~$\eta$ on~$\Mat(\Lambda_D)$ 
and an invertible subalgebra~$\calA$ of~$\Mat(\Lambda_{D-1})$
such that
\begin{equation}
    \zeta = \alpha \circ \beta_\calA \circ \eta
\end{equation}
can be defined on a half space~$\Mat(\Lambda_D^{z<0})$
by modifying its action only within distance~$O(r)$ from the boundary,
while keeping the spread bounded.
From now on we use $z$ as the coordinate along a fixed direction of the lattice.
That is, there exists a QCA~$\zeta'$ on~$\Mat(\Lambda_D^{z<0})$ of spread~$r' = O(r)$ such that
$\zeta(x) = \zeta'(x)$ for all $x \in \Mat(\Lambda_D^{z < -r'}) \subset \Mat(\Lambda_D^{z< 0})$.
This property is called ``blending''~\cite{Freedman2019} or ``crossover''~\cite{Gross2012} to the vacuum.

\begin{figure}[b!]
    \centering
    \includegraphics[width=0.9\textwidth, trim={0ex 47ex 39ex 25ex}, clip]{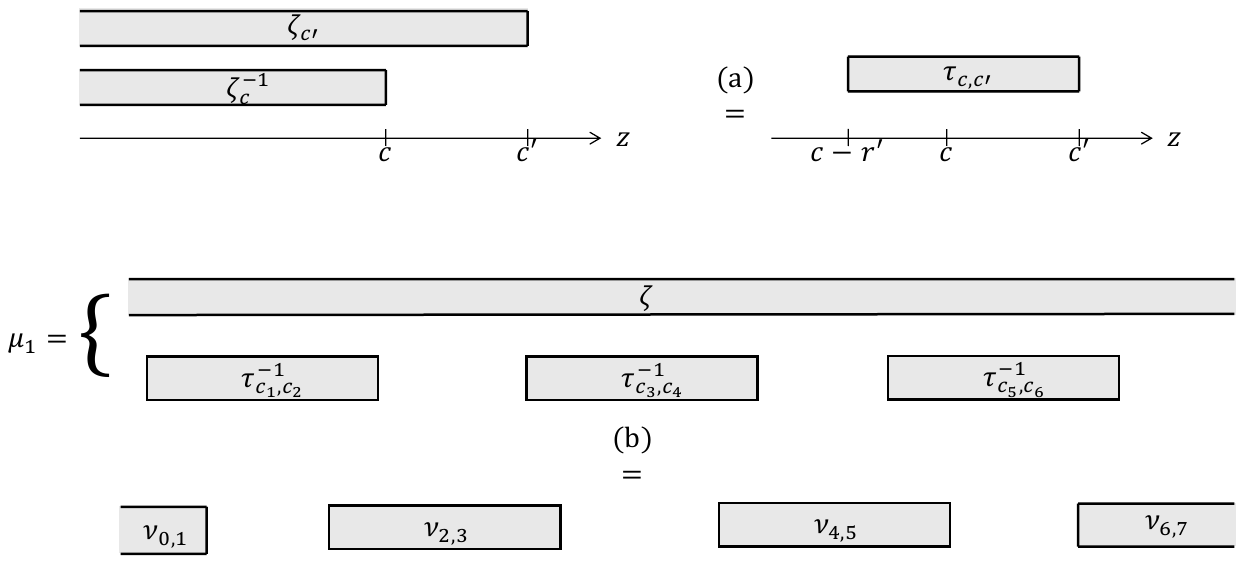}
    \caption{
        The QCA~$\zeta_c$ acts as~$\zeta$ on $z < c - r'$ but as identity on $z \ge c$.
        The composition~$\tau_{c,c'} = \zeta_{c'} \circ \zeta_{c}^{-1}$ 
        is then supported on the region $c - r' \le z < c'$
        which is compact along $z$-direction but otherwise not.
        This is the equality~(a).
        Since $\tau_{c,c'}$ acts by $\zeta$ on the interior of the region~$c - r' \le z < c'$,
        a composition $\mu_1$ of~$\zeta$ and a product of separated $\tau$'s
        gives a product of separated QCA, each of which can be regarded as a QCA in one dimension lower.
        This is the equality~(b).
    }
    \label{fig:blendtocircuit}
\end{figure}

Using the assumption of translation invariance, 
we now show that this blending into the vacuum implies that
the QCA~$\zeta$ is a composition of two layers of nonoverlapping $(D-1)$-dimensional QCA.
The translation invariance implies that $\zeta$ can be defined on a half space 
with a boundary positioned at~$z = c$ for \emph{any}~$c$.
Let $\zeta_c$ be such a QCA on~$\Mat(\Lambda_D^{z < c})$.
By definition, $\zeta_c$ agrees with $\zeta$ on the region where $z < c - r'$.
We trivially extend the domain of definition of~$\zeta_c$ by identity 
so that the extension is defined on the whole~$\Mat(\Lambda_D)$;
the extended $\zeta_c$ acts by the identity for all operators on the region where $z \ge c$.
For $c + 5r' < c'$, we let $\tau_{c,c'} = \zeta_{c'} \circ \zeta_c^{-1}$.
By construction, $\tau_{c,c'}$ acts by identity on the region where $z < c - r'$ or $z \ge c'$,
and acts by $\zeta$ on the region where $c \le z < c' - r'$.
See Figure~\ref{fig:blendtocircuit}.
Now, consider a trivial equation:
\begin{equation}
    \zeta = 
    \underbrace{ \zeta \circ \left[ \prod_{j \in \ZZ} \tau^{-1}_{20 r' j, 20 r' j + 10 r'} \right]}_{\mu_1} 
    \circ 
    \underbrace{\left[ \prod_{j \in \ZZ} \tau_{20 r' j, 20 r' j + 10 r'} \right]}_{\mu_2}
\end{equation}
where the constants $10$ and $20$ are some convenient choices.
Because $\tau^{-1}_{20 r' j, 20 r' j + 10 r'}$ cancels the action of $\zeta$ on the interior of its support,
the first part~$\mu_1$ is a product of QCA supported on disjoint regions.
Hence, both~$\mu_1$ and~$\mu_2$ are each a product of QCA over a collection of separated regions,
each of which can be regarded as a $(D-1)$-dimensional QCA.
By induction hypothesis, $\zeta$ has gate complexity~$O(n)$.
Since we know a circuit for a shift QCA~$\eta$ with gate count~$O(n)$,
the problem of implementing $\alpha$ using a number of gates linear in the system volume 
is now reduced to that of~$\beta_\calA$.

\begin{figure}[b!]
    \centering
    \includegraphics[width=0.8\textwidth, trim={10ex 88ex 73ex 20ex}, clip]{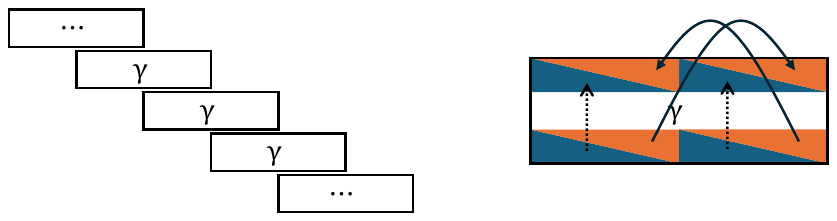}
    \caption{%
        Implementation of~$\beta_\calA$ that pumps an invertible subalgebra~$\calB$ using a swap-like QCA~$\gamma$.
        The blue triangle on the right diagram is~$\calA$ and the orange is~$\calB$.
        By the staircase, only $\calB$ is pumped to the right,
        which is exactly the action of~$\beta_\calA$.
    }
    \label{fig:gamma}
\end{figure}

We show that $\beta_\calA$ can be implemented in a staircase fashion
where each step of the staircase is a $(D-1)$-dimensional QCA~$\gamma$ of a constant spread,
analogous to the swap gate in the $D=1$ case.
Take $k$-th and $(k+1)$-st sheets 
supporting~$\Mat(\Lambda_{D-1})^{(k)} \otimes \Mat(\Lambda_{D-1})^{(k+1)}$.
By definition of the invertible subalgebra~$\calA$,
we have
\begin{equation}
    \Mat(\Lambda_{D-1})^{\otimes 2} \cong \calA^{(k)} \otimes \calB^{(k)} \otimes \calA^{(k+1)} \otimes \calB^{(k+1)} .
\end{equation}
The swap-like QCA~$\gamma$ is defined as the following. See Figure~\ref{fig:gamma}.
\begin{equation}
    \xymatrix{
    \gamma : &\calA^{(k)} \otimes \calB^{(k)} \otimes \calA^{(k+1)} \otimes \calB^{(k+1)} 
    \ar[d]
    &
    \ni
    & 
    a \otimes b \otimes c \otimes d
    \ar@{|-{>}}[d]\\
    &\calA^{(k)} \otimes \calB^{(k)} \otimes \calA^{(k+1)} \otimes \calB^{(k+1)} 
    &  
    \ni
    &
    a \otimes d \otimes c \otimes b
    } 
\end{equation}
Obviously, $\gamma$ is an automorphism of~$\Mat(\Lambda_{D-1})^{\otimes 2}$ with spread given by
a function of~$r$ where $r$ appears in the definition of the invertible subalgebra~$\calA$.
The promised implementation of~$\beta_\calA$ using the QCA~$\gamma$ 
is shown in Figure~\ref{fig:gamma}.
Therefore, the QCA~$\beta_\calA$ is a staircase circuit by the induction hypothesis using $O(n)$ gates.
This completes the proof.

We have converted a QCA~$\alpha$ into a constant number (independent of~$n$) of staircase unitary circuits,
so that every qubit goes through a constant number of local unitary gates.
    We did not answer whether the staircase circuit could be arranged 
so that all the gates were contained in a single hyperplane in spacetime of small thickness.
The final circuit from our inductive proof
consists of a constant number of hyperplanes in spacetime of constant thickness
that contain all the local gates.

In connection to the previous section on the shallow circuit learning problem,
an open problem is whether there is an efficient query algorithm 
for the conversion from a QCA to a staircase circuit.
The two types of data that define a QCA~$\alpha$
--- one a table of images of single-qudit operators under a QCA
and the other a local unitary circuit in a staircase fashion ---
contain the same number $O(n)$ of parameters up to a fixed multiplicative constant.
The former is perhaps the most obvious type of data for~$\alpha$,
but the latter is more appropriate
if we are to apply~$\alpha$ to some arbitrary operator
whose decomposition in terms of single-qudit operators is complicated.

\section*{Acknowledgments}

I thank Hsin-Yuan (Robert) Huang for discussions
and Sumner Hearth, Anushya Chandran, and Chris Laumann
for explaining~\cite{Hearth2023}.

\bibliographystyle{alphaurl}
\bibliography{circuit-refs}

% \onecolumn\newpage
% \appendix

\end{document}